# A framework for analyzing hyper-viscoelastic polymers

A.R. Trivedi & C.R. Siviour
*Solid Mechanics and Materials Engineering Group, Department of Engineering Science, University of Oxford, Parks Road, Oxford, OX1 3PJ, UK*

ABSTRACT: Hyper-viscoelastic polymers have multiple areas of application including aerospace, biomedicine, and automotive. Their mechanical responses are therefore extremely important to understand, particularly because they exhibit strong rate and temperature dependence, including a low temperature brittle transition. Relationships between the response at various strain rates and temperatures are investigated and a framework developed to predict large strain response at rates of c. 1000 s$^{-1}$ and above where experiments are unfeasible. A master curve of the storage modulus's rate dependence at a reference temperature is constructed using a DMA test of the polymer. A frequency sweep spanning two decades and a temperature range from pre-glass transition to pre-melt is used. A fractional derivative model is fitted to the experimental data, and this model's parameters are used to derive stress-strain relationships at a desired strain rate.

## 1 INTRODUCTION

The overall research goal is to be able to predict the mechanical response of polymers and their composites at high strain rates purely by knowing their composition. And vice versa, this would also allow a material to be designed to perform to a certain set of mechanical requirements or constraints. In order to obtain the mechanical properties of polymers at high rates, there are three potential options that were investigated in the current research.

Most simply, to obtain the material response at high strain rates, one could test the polymer at high rates. This would be ideal, however it is not as simple as it initially sounds. Due to the low stiffness of the material, the ensuing low wave speed means that stress equilibrium is not easily reached, and under dynamic loading, the forces measured at the ends of the specimen may not be representative of the material response. This can also be understood as structural vibrations masking the material response as the resonant frequency of the sample is similar to the frequencies in the loading pulse. A second problem, encountered in the split-Hopkinson pressure bar (SHPB), the standard apparatus for material characterization at these rates, is that the forces supported by soft specimens are low, making measurements difficult. Some effort has been made in trying to match bar and specimen impedances to allow the force transmitted to be maximized and this issue to be overcome (Chen 1999, Siviour 2008). A good description of high strain rate testing of low impedance materials is given in Gray & Blumenthal (2000).

The second option allows for a qualitative understanding of the temperature-rate interdependence through a one-to-one equivalence between a characteristic stress at a particular rate and the same characteristic stress at a lower rate, but different temperature. This is based on the classic result of the time-temperature superposition (TTS) principle in which one can quantify changes in rate to changes in temperature (Siviour et al. 2005, Roland 2011, Kendall & Siviour 2014). The issue with this method is that it is only useful for obtaining a descriptive relationship between a particular value of stress and the temperature at which one must conduct the test to obtain similar mechanical properties at a lower strain rate. It is not guaranteeing the full stress-strain relationship would be possible to obtain by this equivalence, and nor is it allowing us to predict the mechanical response since it refers to experimental data points that must have been collected prior to conducting this time-temperature equivalence.

The third option is where the majority of the work done in the current research is concentrated. It relies on a dynamic mechanical and thermal analysis

(DMA) test and the ability to use TTS to obtain master curves and shift factors detailing the material's rate and temperature dependences. By fitting a suitable model to the experimental data from the DMA test, it is possible to obtain material parameters for a constitutive model that would allow for the prediction of the stress-strain response at any appropriate strain rate.

## 2 EXPERIMENTAL OVERVIEW

### 2.1 Material

The polymer material used for the current research as presented in this paper was a commercial black neoprene rubber supplied in the form of a 5 mm sheet by Brammer, UK.

### 2.2 Dynamic mechanical and thermal analysis

For the DMA tests, rectangular samples of the material were used with dimensions $50 \times 10 \times 5$ mm. The tests were conducted on a TA Instruments Q800 and all tests were performed in the dual cantilever configuration. An isothermal frequency sweep in 2 °C increments was used with values of 0.5, 2, 5, and 10 Hz and a temperature range from $-50$ °C to $80$ °C.

### 2.3 Compression tests

From the neoprene rubber sheet, cylindrical samples with a diameter of 5 mm were cut out for use in the compression tests. Compression tests were conducted at a variety of strain rates from $10^{-3}$ to $2100$ s$^{-1}$ at a temperature of 25 °C, and at a variety of temperatures from $-100$ °C to 80 °C at a rate of $10^{-2}$ s$^{-1}$.

#### 2.3.1 Low strain rate compression tests

A commercially available Instron 5980 electromechanical static testing machine was used for tests at strain rates of $10^{-3}$, $10^{-2}$, and $10^{-1}$ s$^{-1}$. A 5 kN load cell was used for measurements and the machine crosshead displacement to obtain strain. Preliminary experiments showed that at room temperature there was minimal machine compliance, so the cross-head displacement gave an accurate measure of the movement of the loading platens. Constant time strain rate control was used in all experiments.

When testing at a rate of $10^{-2}$ s$^{-1}$ at sub-ambient temperatures, an environmental chamber was used with a liquid nitrogen feed. Temperature was controlled with the default thermostat which was accurate to $\pm 1$ °C. The environmental chamber's feedback control thermocouple is situated at the back of the chamber away from the specimen, and so a second thermocouple was inserted into the loading platen itself to measure the temperature as close to the specimen as possible. Only when this thermocouple reading matched the chamber's thermocouple, was the test conducted. The same chamber and temperature measurements were used for supra-ambient temperature tests.

At lower temperatures, due to higher loads experienced, there was a need to increase the load cell to 50 kN and as the higher loads could lead to rig compliance, an extensometer with a sufficiently large gauge length of 12 mm was used to obtain local strain measurements.

#### 2.3.2 High strain rate compression tests

For compression tests at a strain rate of the order $10^3$ s$^{-1}$, an in-house SHPB was used. Details of the analysis procedure can be found in Gray & Blumenthal (2000).

Figure 1 shows a schematic of the SHPB set-up. Ti-6Al-4V alloy bars with a 12.7 mm diameter were used. Using compressed gas to fire the striker bar, speeds of up to around 20 m s$^{-1}$ were achievable giving the possibility of testing at rates of order $10^3$ s$^{-1}$.

## 3 RESULTS AND ANALYSIS

### 3.1 Varying rate tests

The results of the low strain rate tests on the rubber sample are presented in Figure 2. It is evident from the results of the low strain rate tests that when the strain rate increases, the stress experienced at any level of strain is higher. If we consider a characteristic stress at a strain value of 0.1, then the stress is seen to increase from 0.79 MPa at a rate of $10^{-3}$ s$^{-1}$ to 1.00 MPa at a rate of $10^{-1}$ s$^{-1}$. This represents on average a 13 % increase in characteristic stress per decade at these low strain rates.

Results of the high strain rate tests are presented in Figure 3. If we again consider the characteristic stress at a strain value of 0.1, we find that average at a strain rate order of $10^{-3}$ s$^{-1}$ is 6.5 MPa. This is a significant increase from the lower orders of strain rate. The oscillations on the stress-strain curves are a result of structural vibrations in the specimen. These would be more significant for lower modulus specimens or higher strain rates.

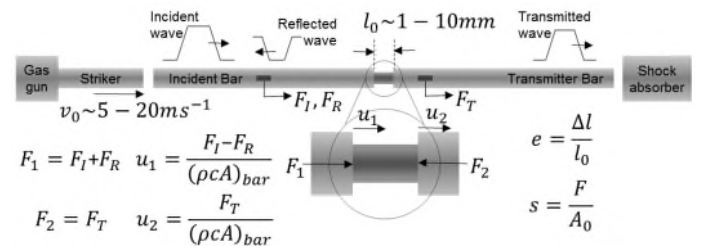

Figure 1. Schematic of the SHPB set-up used for the high strain rate tests.

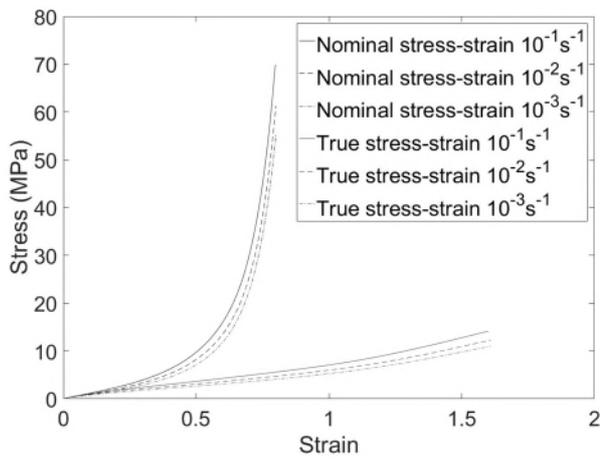

Figure 2. Comparison of stress-strain relationships for the rubber sample averaged over a series of three tests conducted at a variety of low strain rates.

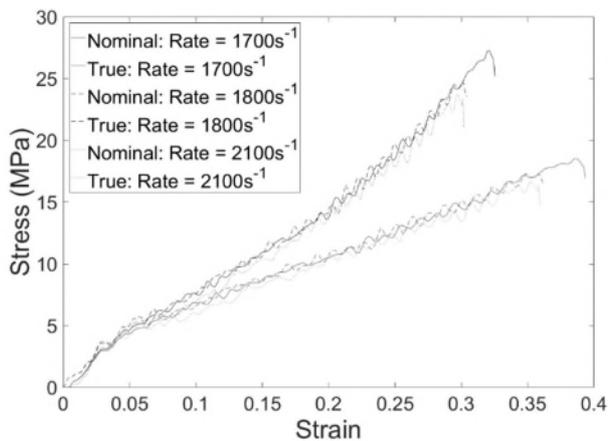

Figure 3. Comparison of stress-strain relationships for the rubber sample over a series of three tests conducted at a variety of high strain rates.

The overall behavior can be characterized by the graph in Figure 4. Although further data are required to fully define the rate dependence, from previous studies we expect an approximately bilinear relationship between stress and strain rate, with a stronger dependence in the high rate experiments. The effect of glass transition leading to the rate dependence of the mechanical properties of the rubber is clear to see, and will now be demonstrated by comparison to tests at different temperatures.

### 3.2 Varying temperature tests

The results of the compression tests conducted at a variety of temperatures are presented in Figure 5.

From these uniaxial tests, it is clear to see that there is a remarkable distinction in mechanical properties at lower temperatures compared to higher temperatures. The transition occurs just below −40 °C. Above this temperature, the response of the material is rubbery. Below this temperature, the response is glassy and shows brittle fracture characteristics. In fact, this was noticeable from physical observations of the rubber post testing.

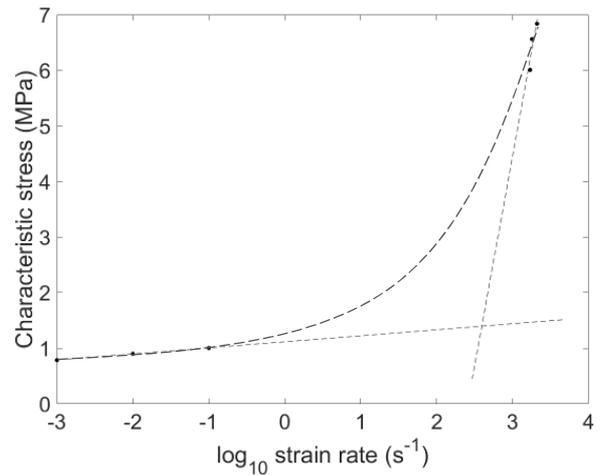

Figure 4. Behavior of characteristic stress as a function of the strain rate.

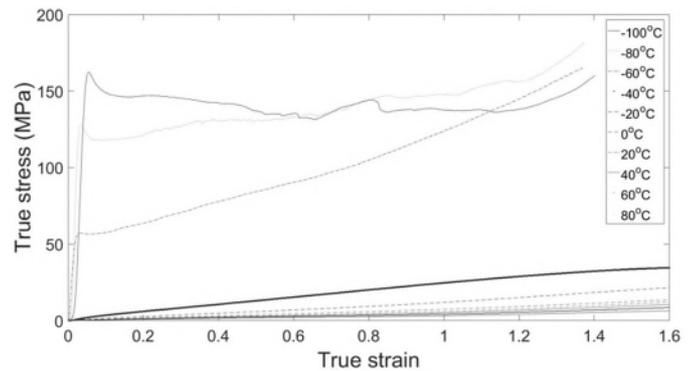

Figure 5. Comparison of stress-strain relationships for the rubber sample averaged over a series of three tests conducted at a variety of temperatures at a strain rate of $10^{-2}$ s$^{-1}$.

Figure 6 shows a similar dependence on temperature as Figure 4 showed with the strain rate.

### 3.3 Temperature-rate equivalence

Qualitatively, it is already possible to note using Figures 4, 6 that high stresses occur at both high strain rates and low temperatures. So it does not seem too ridiculous to postulate that there may be some sort of equivalence in material response between high rate compression tests and low temperature compression tests conducted at a lower strain rate.

For example, if we consider where we want to know the likely stress-strain relationship of this rubber at a strain rate of 2000 s$^{-1}$, using Figure 4, we note that the characteristic stress at this strain rate is around 6.5 MPa. This same value of characteristic stress is present in Figure 6 at a temperature of − 45 °C. This means we can obtain a representative stress-strain relationship for the rubber by using this equivalence and testing it under compression at − 45 °C and a strain rate of $10^{-2}$ s$^{-1}$.

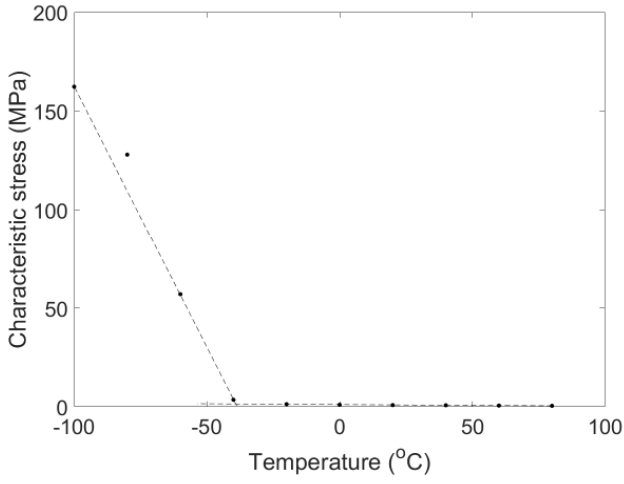

Figure 6. Behavior of characteristic stress as a function of the temperature.

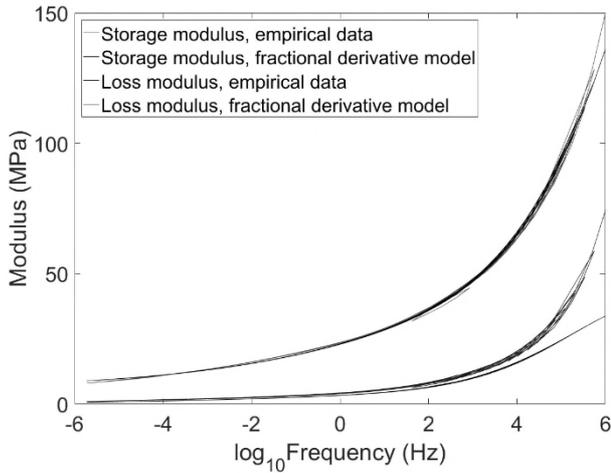

Figure 7. Master curve for averaged DMA data at a reference temperature of 25 °C.

## 3.4 DMA tests

A DMA test was conducted as explained in §2.2, and TTS was performed with the data by shifting data at various temperature steps parallel to the logarithmic frequency axis by a certain shift factor to create a master curve from this series of overlapping data.

Once this has been completed with all the temperature steps, the master curve as shown in Figure 7 is obtained. Further details of how to construct master curves can be found in literature (Morrison 2001, Tobolsky 1960).

## 4 MODELLING AND SIMULATIONS

### 4.1 Hyperelastic constitutive model

To model the hyperelastic behavior of the rubber, an Ogden model (1972, 1984) was used as it has a well-established base in literature (Shergold et al. 2006, Yoon 2016, Kossa & Berezvai 2016). It is based on isotropic, isothermal, and incompressible assumptions. For the case of the rubber in this paper, these assumptions are valid.

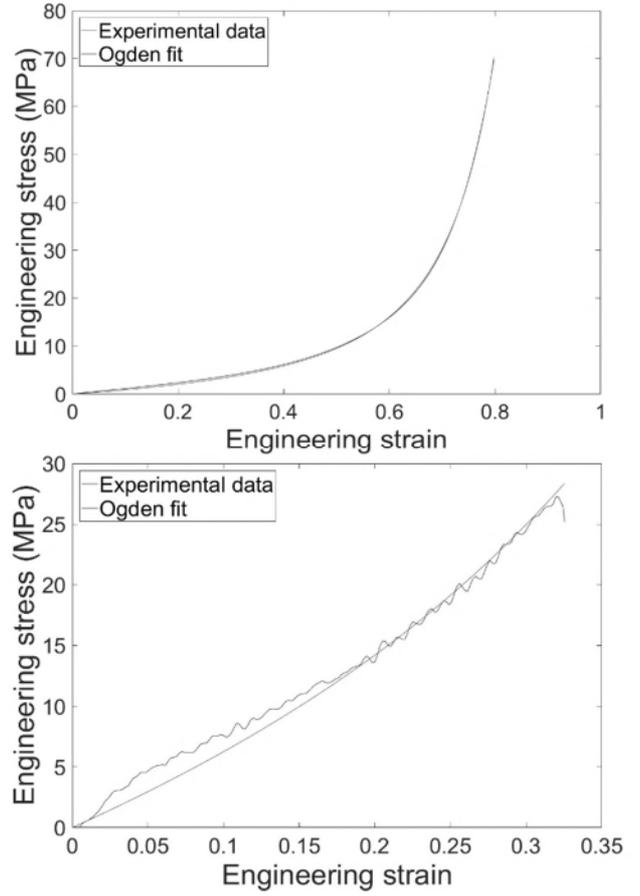

Figure 8. Ogden model fit for samples tested at a rate of $10^{-1}$ s$^{-1}$ (a, top) and one tested at a rate of 2100 s$^{-1}$ (b, bottom).

$$\phi = \frac{2\mu}{\alpha^2}(\lambda_1^\alpha + \lambda_2^\alpha + \lambda_3^\alpha - 3) \quad (1)$$

$$\sigma_3 = \frac{2\mu}{\alpha}\left[\lambda_3^{\alpha-1} - \lambda_3^{-(1+\alpha/2)}\right] \quad (2)$$

The Ogden model in Equation 1 is based on $\phi$, a strain energy density per unit volume; $\mu$ = representative shear modulus; $\alpha$ = strain hardening exponent; and $\lambda_i$ = principle stretch ratios for the three Cartesian directions. Based on the special case of the Ogden where the material is incompressible and exposed to uniaxial stress loading, it is possible to obtain the stress relationship as shown in Equation 2.

In Figure 8a, the Ogden model applied using a rate independent strain hardening exponent and a rate dependent shear modulus term shows an excellent fit to the experimental data. In figure 8b, the fit is still very good but deviates from the experiment at low levels of strain. This is due to the added rate dependence of the viscoelastic component to the stress at the higher rate.

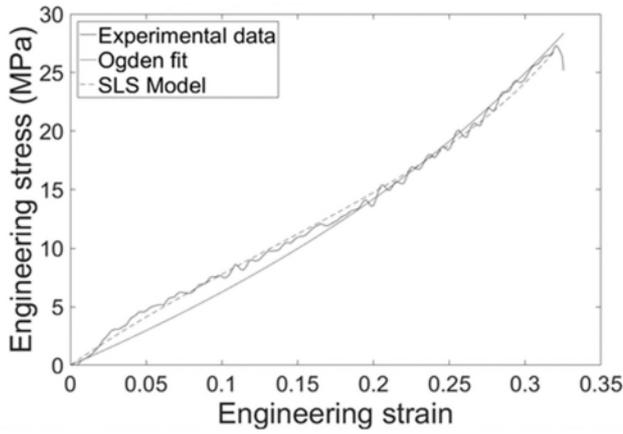

Figure 9. Comparison of the fits to the experimental data using either a hyperelastic Ogden model or an SLS model.

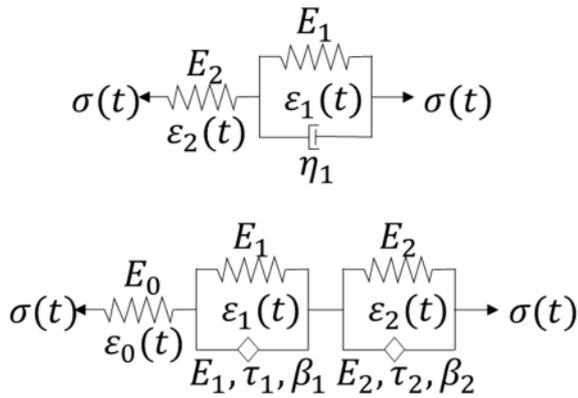

Figure 10. Diagrammatic representation of the SLS model (a, top) and the fractional SLS model (b, bottom).

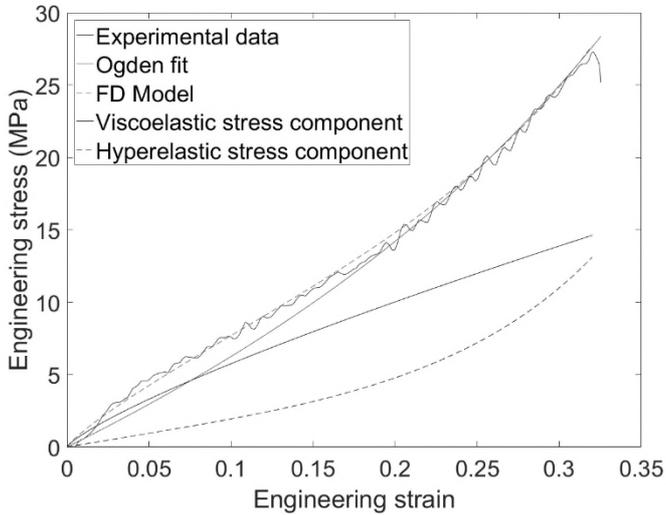

Figure 11: Comparison of the fits to the experimental data using the hyperelastic Ogden model as seen before, and the result of the fractional derivative model, along with its components.

### 4.2 Viscoelastic constitutive model

To capture the low strain viscosity that was lacking in the hyperelastic model in Figure 8b, the next step of the modelling process was to decompose the stresses into their hyperelastic and viscoelastic components. The hyperelastic element was the same as the Ogden model in §4.1, but a Standard Linear Solid (SLS) model was used for the viscoelastic component. A diagrammatic representation of which can be found in Figure 10a. This decomposition is seen in recent literature like Bai (2016) and is well suited to model the rubber here.

Figure 9 shows the result of the superposition of the hyperelastic stress and viscoelastic stress components to form the SLS curve which clearly follows the experimental data much better than the simple Ogden model.

### 4.3 Fractional derivative constitutive model

Although it has been possible to use constitutive models to provide excellent fits to the experimental data, the aim of the current research is to be able to predict the high strain rate response of the rubber doing a minimum number of experimental tests, especially those conducted at an unfeasible higher strain rate. In order to achieve this, a fractional derivative constitutive model is fitted directly to the DMA results; the results can be seen again in Figure 7.

In modelling the storage modulus, the fractional SLS model shown in Figure 10b provides an excellent fit. There is also a decent fit at lower frequencies for the loss modulus. The storage and loss moduli are obtained by solving the differential equation for the fractional SLS model to obtain a complex modulus as shown in Equation 3 and then taking either the real part (storage modulus) or the imaginary part (loss modulus).

In order to obtain a stress-strain relationship, the modulus in the time domain needs to be found first. To convert the complex modulus into time domain is quite cumbersome when the values of $\beta$ are not equal and take the value of one (reduces the problem to that of the SLS leading to an exponential time domain term) or a half (leads to an error function term in time domain as seen in Koeller (2007)). Hence, an approximate interconversion method is used with Equations 4-6, which has been verified with examples (Schapery & Park 1999).

Once the modulus in time domain is found, a standard hereditary integral operation using Equation 7 can be performed of which examples can be found in Park (2001) and Bai (2016). Since the majority of points from the results of the TTS are at low values of time, it makes sense to change variables such that the integral can be performed using $\ln t$ instead as shown in Equation 8. To counter the issue with the limit of the integral starting from $\ln 0$, an approximate initial value of stress $\sigma_0 = E_0 \dot{\varepsilon} e^{du}$ is used instead, where $E_0$ is the instantaneous modulus, $\dot{\varepsilon}$ the constant strain rate, and $du$ the first increment in ln-time.

The viscoelastic stress obtained in this manner can then be superposed with the hyperelastic component from the Ogden model described in §4.1 with the rate-dependent terms replaced with independent ones. The results of this superposition can be seen in Figure 11.

$$E^* = E_0 + \frac{E_1(i\omega)^{\beta_1}}{(i\omega)^{\beta_1} + 1/\tau_1^{\beta_1}} + \frac{E_2(i\omega)^{\beta_2}}{(i\omega)^{\beta_2} + 1/\tau_2^{\beta_2}} \quad (3)$$

$$E(t) \cong \frac{1}{\tilde{\lambda}} \tilde{E}(s)\Big|_{s=1/t} \quad (4)$$

$$\tilde{\lambda} = \Gamma(1-n) \quad (5)$$

$$n = \frac{d \log \tilde{E}(s)}{d \log s} \quad (6)$$

$$\sigma = \int_0^t E(t-\xi) \frac{d\varepsilon}{d\xi} d\xi \quad (7)$$

$$\sigma = \dot{\varepsilon} \int_{\ln 0}^{\ln t} E(t - e^u) e^u du \quad (8)$$

## 5 CONCLUSIONS

This paper presents what has been possible so far in providing a framework for better understanding mechanical properties of polymers at high strain rates. This has been done mainly by fitting a fractional constitutive model to DMA data, and then using that model to verify the stress-strain relationship at a strain rate of 2100 s$^{-1}$. However, there are some challenges which allows for further research to be conducted in this area.

The predictive intention of the framework needs to be fully verified with the investigation of previously unexplored strain rates; finding the stress-strain relationship from the model before doing the actual experiment to see whether it is indeed correct as expected. An FEA model will also be produced which will be able to simulate the experimental behaviour and can be used to verify the predictive framework.

Overcoming these obstacles will allow for a truly cohesive framework to be developed to allow for the prediction of high strain rate properties of hyper-viscoelastic polymers and their composites.

## 6 ACKNOWLEDGEMENTS

This material is based upon work supported by the Air Force Office of Scientific Research, Air Force Materiel Command, USAF under Award No. FA9550-15-1-0448. Any opinions, findings, and conclusions or recommendations expressed in this publication are those of the author(s) and do not necessarily reflect the views of the Air Force Office of Scientific Research, Air Force Materiel Command, USAF.

This author would also like to thank Richard Duffin of the Engineering Science Solid Mechanics workshop for his technical assistance, and Nick Hawkins of the Oxford Silk Group in the Zoology Department, University of Oxford for his ongoing support with the TA Q800 DMA apparatus.